\documentstyle[epsf]{article}
\newcommand{\bee}{\begin{equation}}
\newcommand{\ee}{\end{equation}}
\newcommand{\beea}{\begin{eqnarray}}
\newcommand{\eea}{\end{eqnarray}}

\newcommand{\ewxy}[2]{\setlength{\epsfxsize}{#2}\epsfbox[10 60 640 570]{#1}}

\begin{document}
\thispagestyle{empty}
\parskip=12pt
\raggedbottom

\def\mytoday#1{{ } \ifcase\month \or
 January\or February\or March\or April\or May\or June\or
 July\or August\or September\or October\or November\or December\fi
%\space\number\day ,
 \space \number\year}
\hspace*{9cm} COLO-HEP-391\\
\vspace*{1cm}
\begin{center}
{\LARGE Instanton Effects in  Hadron Spectroscopy in SU(2) (Lattice)
 Gauge Theory}
\vspace{0.5cm}

Thomas DeGrand,
Anna Hasenfratz,  and Tam\'as G.\ Kov\'acs\\
Department of Physics \\
 University of Colorado,
Boulder CO 80309-390 

\vspace{1cm}

\mytoday \\ \vspace*{1cm}

\nopagebreak[4]
\vspace{.5cm}
\begin{abstract}
We describe  quenched spectroscopy in $SU(2)$
gauge theory using 
smoothed gauge field configurations. We investigate the properties
of quarks moving in instanton background field configurations, where the
sizes and locations of the instantons are taken from simulations
of the full gauge theory.
By themselves, these multi-instanton configurations do not confine quarks, but
they induce chiral symmetry breaking.
\end{abstract}
\end{center}

\section{Introduction}
What features of the QCD vacuum are responsible for confinement or
for the generation of the observed structure of hadron spectroscopy?
  This question might,
in principle, be answered by lattice simulations of non-Abelian gauge
theories.
In practice, however, the dominant features of the QCD vacuum as seen
in lattice simulations are basically identical to the dominant features of
the vacuum of
any field theory, be it confining or not: short distance fluctuations of
the dynamical variables, whose main effect is simply to renormalize the
bare parameters of the theory.  These fluctuations
mask the (presumably) longer-distance features which are the true objects
of the question.

This complication has a standard solution in lattice simulations, 
namely to transform the  lattice
in a way that reduces
short-distance fluctuations,
but leaves the long distance properties unchanged.
Observables constructed from the new variables correspond to extended,
sometimes non-local observables on the original lattice.
Examples of these constructions include APE-blocked links \cite{APEBlock}, used
in measurements of the glueball spectrum and string tension, 
``cooled links''\cite{COOL}
 used in many investigations of topology on the lattice,
or the use of ``smeared operators'' \cite{SMEAR}
in matrix element calculations.
The separation of structure in the vacuum into two classes,
short distance vs. long distance, is intrinsically ambiguous, and this
can contaminate one's results. For example, instantons are localized objects of
size about 0.2-0.3 $fm$, i.e. they are not long distance quantities in the
continuum. On the lattice at small lattice spacing they span large distance in
lattice units and a local transformation will preserve them. On coarse lattices,
however, their size in lattice units is small and a 
smoothing transformation might
lose physically relevant objects. 
If one wants to study the vacuum by using particular ultraviolet-insensitive
observables/transformation, one ought to be certain that they do not
alter the long distance behavior one is interested in.

Once  a particular lattice transformation reveals the structure of the vacuum, we
can ask in what way a particular set of dynamical degrees of freedom is
responsible for the dynamics of the system. We  would regard the evidence for the
importance convincing only,
if an approximation to the vacuum which included only the considered degrees of
freedom predicted hadronic properties in rough quantitative
agreement with those from the full theory.

In this note we concentrate on instanton excitations of the vacuum.
Based on phenomenological models, it has been argued that instantons
are largely responsible for chiral symmetry breaking and
the low energy hadron and glueball spectrum
\cite{Diakonov,Shuryak_long}. Instanton liquid
models attempt to reproduce the topological content of the QCD
vacuum and conclude that hadronic correlators in the instanton liquid
show all the important properties of the corresponding full
QCD correlators. These models appear to capture the essence of
the QCD vacuum, but their derivations involve a number of uncontrolled
approximations and phenomenological parameters.

We are motivated to study this problem because of our work
on instantons \cite{INST}, where we devised a method of constructing smoothed
gauge configurations
whose long distance behavior was identical to the original 
configurations, but which had tiny short distance fluctuations. 
 We found that
most of the action of the smoothed configurations is carried
by instantons. But 
to what extent do instantons affect long distance physics such
as confinement or hadron structure? We use the smoothed configurations
as templates to construct configurations containing only instantons,
and then investigate the spectral properties of quarks moving in these
background configurations.

Our smoothing mechanism is called ``cycling.''  
Imagine beginning with a set of lattice variables $\{V\}$ 
on a lattice whose spacing is $a$ and lattice size is $L$.
The lattice action is a fixed-point action \cite{FP} $S^{FP}(U)$.
The first part of a cycling step is 
 done by performing
an inverse blocking transformation
 to construct a set of fine lattice variables
$\{U\}$ occupying a lattice of lattice spacing $a/2$ and
lattice size $2L$, by solving the
steepest-descent equation
\bee
S^{FP}(V)=\min_{ \{U\} } \left( S^{FP}(U) +\kappa T(U,V)\right) . \label{STEEP}
\ee
Here $\kappa T(U,V)$ is the renormalization group blocking kernel.
Now the original lattice occupies one of the 16 sublattices of
the fine lattice.  Performing an RG blocking transformation from the
fine variables to a set of variables based on the sublattice 
corresponding to the original lattice merely inverts the minimization of
Eq. \ref{STEEP}.
However, if we perform a blocking transformation to a set of coarse variables
 $\{W\}$ based on one of the other sublattices, the delicate coherence
among the fine variables
will be broken and the new coarse variables 
will be strongly ordered on the
shortest distance scale (as measured, for example, by the expectation value
of the plaquette) while retaining all long distance physics (because they
are generated by a RG blocking transformation).
This is the second part of the cycling transformation 
$V_\mu(x)\rightarrow U_\mu(x) \rightarrow W_\mu(x)$.
Cycling steps can be iterated, and a few cycling steps can reduce
the plaquette to within 0.001 of its free-field value.

One can think of performing spectroscopy on cycled 
gauge configurations as performing spectroscopy on the original 
configurations, but using a more complicated fermion action in which
the naive gauge connections are replaced by cycled links.
This is similar in spirit to the use of ``fat links'' by
the MILC collaboration \cite{MILCIMPR} or the complicated gauge connections
of the approximate FP actions developed by one of us \cite{TOM}.
The ``cycled fermions'' are more insensitive to short distance fluctuations
of the gauge fields than the original fermions.

By itself, cycling  should not be expected to dramatically
improve scaling, since the discretization errors of the
underlying free fermion action are unchanged.
However, cycling  decouples the fermions from short distance
fluctuations present in the simulations.  (This is in contrast to 
tadpole improvement \cite{TADPOLE}, 
as it is usually implemented in simulations, 
in which couplings are merely rescaled.) 
For Wilson fermions we expect to see the critical hopping parameter
$\kappa_c$ closer to 1/8, even at strong coupling.
For staggered fermions, we expect to see a dramatic improvement in
flavor symmetry restoration, as given by the mass difference between the 
Goldstone and non-Goldstone local pions.

The simulations reported here were performed in $SU(2)$ gauge theory
using the FP gauge action of Ref. \cite{INST}. We worked at a coupling
 $\beta=1.5$ (close to the critical coupling for deconfinement at 
$N_t=4$ with  string tension $a^2\sigma=0.122(10)$
and  Sommer radius $r_0/a=3.48(2)$).
We cycled up to 9 times
a set of about 30-40 configurations of size $8^3\times 16$ sites.
We computed Wilson spectroscopy on one source time slice, and staggered
spectroscopy on two time slices, per lattice.
The staggered fermions have antiperiodic temporal boundary conditions;
the Wilson fermions used periodic boundary conditions.
We only performed meson spectroscopy.
In this note we concentrate on staggered fermions, since they provide
a cleaner signal of the physics of interest.

\begin{figure}
\centerline{\ewxy{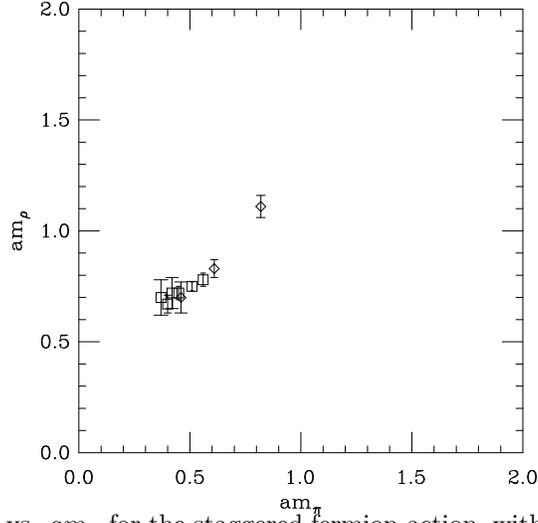}{80mm}
}
\caption{$am_\rho$ vs. $am_\pi$ for the staggered fermion action,
with untouched gauge links (diamonds) and 9-cycled links (squares).}
\label{fig:rhopik}
\end{figure}

Fig. \ref{fig:rhopik} shows $am_\rho$ vs. $am_\pi$ for the 
staggered fermion action,
with raw gauge links and 9-cycled links. 
It appears that cycling does not affect this sector of spectroscopy.

\begin{figure}
\centerline{\ewxy{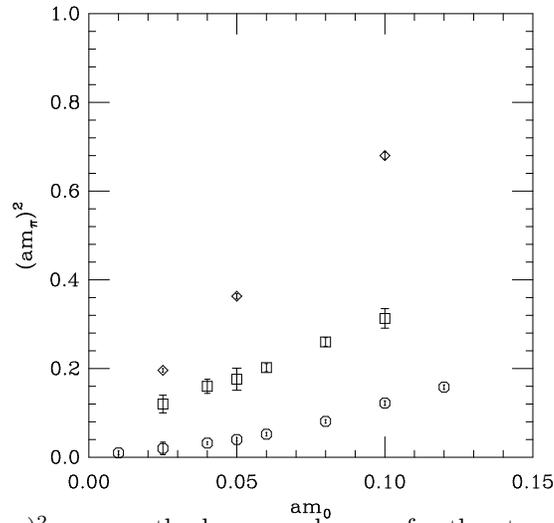}{80mm}
}
\caption{$(am_\pi)^2$ vs. $m_0$ the bare quark mass
for the staggered action,
with raw gauge links (diamonds) and 9-cycled links (squares).
The lightest mass in the pseudoscalar channel with instanton background
configurations is shown by octagons.}
\label{fig:pi20k}
\end{figure}

\begin{figure}
\centerline{\ewxy{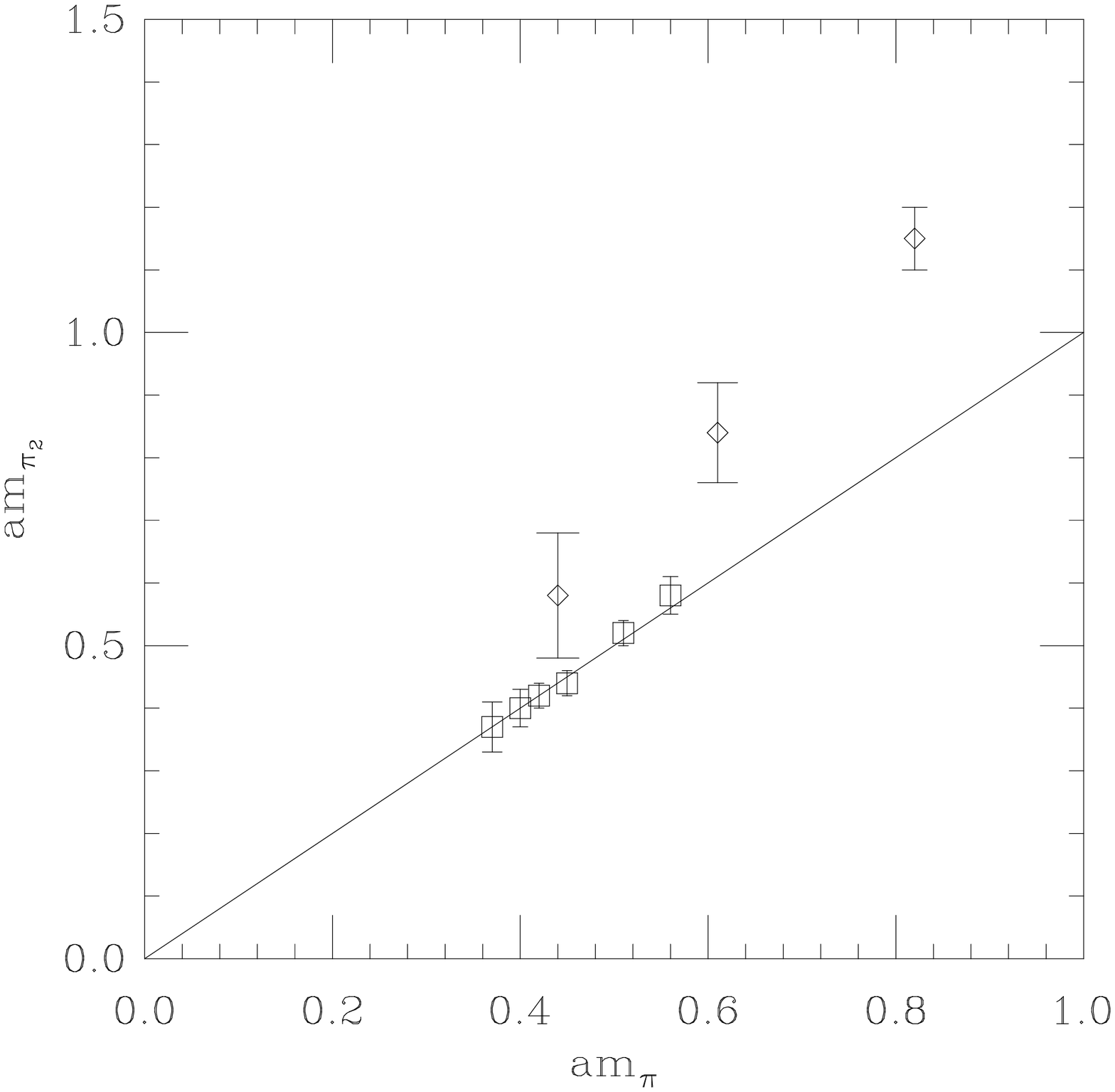}{80mm}
}
\caption{$am_{\pi_2}$ vs. $am_\pi$ for the staggered fermion action,
with raw gauge links (diamonds) and  9-cycled links (squares).
The straight line is $am_{\pi_2} = am_\pi$.}
\label{fig:pi2pi2}
\end{figure}

However, the picture is quite different for the non-Goldstone
partner of the pion, as shown in Fig. \ref{fig:pi20k} and
\ref{fig:pi2pi2}. On the original 
lattices the two mesons have a mass ratio of about 1.4.
On the 9-cycled lattices the two pions are degenerate within
observational uncertainty.

Note that in Fig. \ref{fig:pi20k} the bare mass needed
to produce the same pion mass is much larger on the cycled lattice
than on the original lattice.
This is quite reasonable given the
observed size of fluctuations on the shortest distance scale:
put differently, $f_\pi^2 m_\pi^2 = Z m_q \langle \bar \psi \psi \rangle$
with $Z$ much smaller on the smoothed lattices than on the original ones.

All of these results are anticipated by the ``fat link'' action of the
MILC collaboration\cite{MILCIMPR}.  
There, work is done in $SU(3)$ at $\beta=5.85$,
closer to $\beta_c(N_t=5-6)$.
The pions in that work begin with $m_{\pi 2}/m_\pi \simeq 1.6$,
and their smoothing reduces this number to about 1.2.
Similar results, using a more complicated blocking procedure, have
been reported by  J.-F. Laga\"e and D.~K. Sinclair \cite{SINCLAIR}.

The purpose of this exercise is not to test spectroscopy in detail.
The important point is that the
 action of the many-times-cycled configurations is dominated by
instantons, and yet   the heavy quark
potential and hadron spectroscopy 
are basically unchanged from what we saw on untouched configurations. 

We believe that the physical picture we see at this point is very similar to 
one recently presented by Ivanenko and Negele \cite{NEGELE}.
These authors calculated the lowest eigenvalues of the Dirac operator
and found that the rho and pion contributions to the vector
and pseudoscalar correlation functions were dominated by these 
lowest eigenfunctions.
The lowest eigenfunctions are sensitive to the positions of the instantons
in the background gauge fields.

But these results do not answer the question:
 Are instantons responsible for long distance physics,
or are the structures responsible for long distance physics some other
objects carrying low action, which have been preserved by the cycling transformation?

We can test that question by preparing gauge configurations containing
only instantons and anti-instantons, and using those configurations
as an ensemble for spectroscopy calculations.
We do this by identifying the instantons' sizes and locations 
from many-times cycled configurations, then constructing gauge field
configurations appropriate to the superposition of instantons.
The instanton identification is always done on the inverse blocked
fine lattices, where the positions and sizes of instantons are
measured. Then the instanton content is reconstructed on the same
size lattice in the following way.

Our starting point is a set of lattice configurations
containing single instantons with radii 1.0, 1.5, 2.0...7.0 lattice
spacings. These are
obtained by discretizing continuum instantons in the singular gauge
(see Refs. \cite{putinst}). In this gauge the vector potential
falls off sufficiently rapidly so that in our ensemble the instantons
do not overlap significantly and any sensible ansatz can be used to
combine their fields.
For each instanton found in a given 9-cycled configuration
we took the one in our sequence that was closest in size
and shifted it to the appropriate position. The vector potential
of single instantons was approximated by the logarithm of the links.
We added the vector potentials of all the instantons found in the given
configuration and finally reexponentiated the sum to get the link
SU(2) elements. After the instanton content was reconstructed on the
fine lattice, it was blocked and all the measurements were
performed on the blocked ``coarse'' lattices.
 
Since we have not determined the relative orientation of instantons in
group space, in our earlier work we constructed the artificial
instanton configurations from ``parallel'' instantons. In the present
work we also built instanton configurations where the instanton
orientations are randomly distributed according to the SU(2) Haar measure.
This was done by performing a random global (constant in space-time) gauge
transformation on each individual lattice instanton before using it to construct
the given instanton configuration.

We  now compare the three ensembles; the 9 times cycled real configurations,
the parallel and the randomly oriented instanton configurations with the 
instanton sizes and locations exactly reproduced from the 9-cycled configurations.
It is interesting to note that the average action per configuration is
11.9 on the 9-cycled, 7.9 on the randomly and 8.4 on the parallelly
oriented instanton ensemble (in units of the one-instanton action).
The average number of instantons per configuration is 6.7.  

What is the scale of these ensembles? The 9 times cycled configuration has the
same lattice spacing as the original lattice by construction.
One might want to argue that the physical scale  is
different for the instanton-background simulations.
 In instanton liquid models, the
physical scale is set by the size and/or separation of the instantons,
which in these simulations is derived from the instanton sizes and
locations on the original gauge configurations. So we believe that the
most natural choice for a physical scale in these configurations is
the same as on the original configurations.

\begin{figure}
\centerline{\ewxy{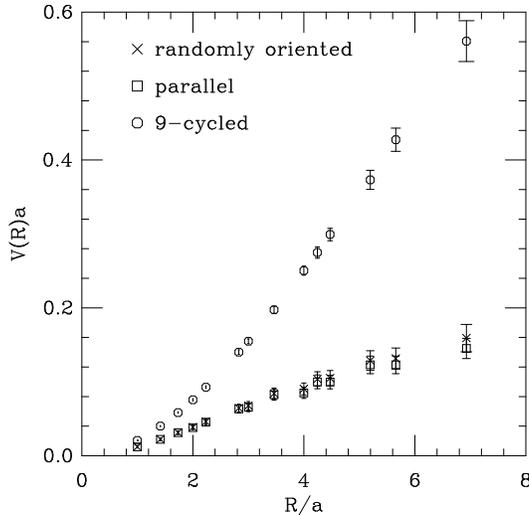}{80mm}}
\caption{The heavy-quark potential measured on the 9 times cycled
real configurations (octogons), the randomly (crosses) and the parallelly
(squares) oriented instantons.}
\label{fig:pot_inst_all}
\end{figure}

In Fig.\ \ref{fig:pot_inst_all} the heavy-quark potentials obtained from the 
three ensembles are compared. We can conclude that neither the parallel
nor the randomly oriented instantons confine. In fact the potentials
produced by the parallel and randomly oriented instanton ensembles are 
hardly distinguishable. It is true that in the real configurations instantons
are probably neither parallel nor completely randomly oriented but since 
the difference between these two extremes is so small, it is  unlikely
that any special orientation arrangement can produce substantially different
results. It seems that instantons by themselves are
not responsible for confinement.

We have also computed spectroscopy on the two types of
instanton ensembles using both  staggered and Wilson fermions.
The dominant feature of both spectroscopy calculation is that
the quarks are deconfined. This is seen most easily in
the staggered fermion spectroscopy. 

\begin{figure}
\centerline{\ewxy{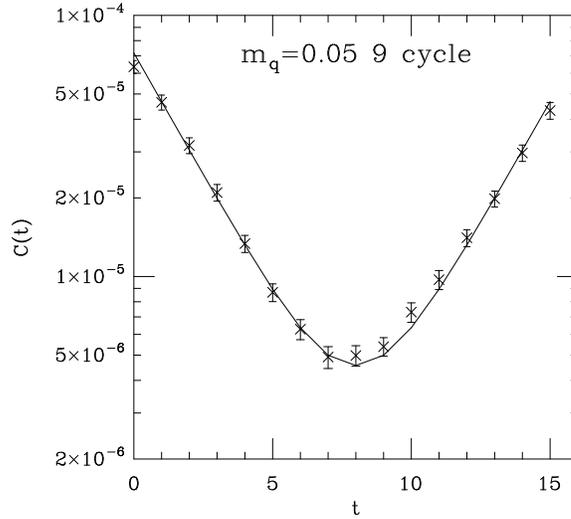}{80mm}
}
\caption{The pseudoscalar propagator from 9-cycled configurations, with
staggered fermions of bare mass $am_0=0.05$.}
\label{fig:pseudo905}
\end{figure}

\begin{figure}
\centerline{\ewxy{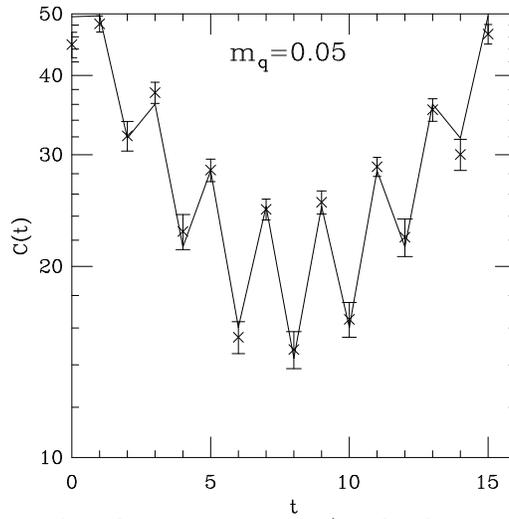}{80mm}
}
\caption{The pseudoscalar propagator in (randomly rotated)
 instanton background
 configurations, with
staggered fermions of bare mass $am_0=0.05$.
The curve is a fit to a single propagating particle plus the
$q \bar q$ branch cut.}
\label{fig:pseudo05}
\end{figure}

Consider the pseudoscalar propagator of the 9-cycled configurations,
shown for one quark mass in Fig. \ref{fig:pseudo905}. It looks like 
any generic lattice pseudoscalar, a more-or-less pure hyperbolic
cosine with no oscillations.
The staggered fermion pseudoscalar propagators on instanton background fields
are quite different: they show the characteristic
sawtooth pattern of free antiperiodic  fermions. This is shown in Fig.
 \ref{fig:pseudo05}.

For Wilson fermions, 
the correlators are qualitatively similar to the correlators of free
$q \bar q$ pairs on the instanton configurations.
A single-exponential fit to the pseudoscalar 
and vector channels gives  masses which fall to zero as the hopping parameter
approaches 1/8 and the   vector and pseudoscalar remain degenerate.

\begin{figure}
\centerline{\ewxy{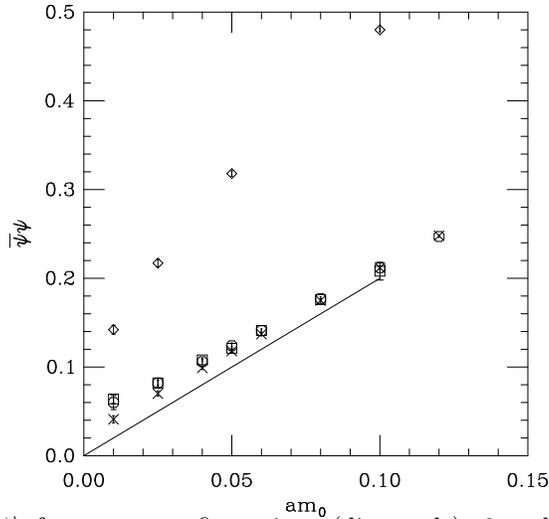}{80mm}
}
\caption{$\langle \bar \psi \psi \rangle$ from raw configurations (diamonds),
 9-cycled configurations (squares), and  instanton-background configurations
which are parallel (crosses) and randomly oriented (octagons),
vs. bare quark mass $am_0$. The line shows the free-field value, $2m_0$.}
\label{fig:pbp}
\end{figure}

And yet, there is physics beyond deconfinement
in these configurations. Fig.  \ref{fig:pbp} shows 
$\langle \bar \psi \psi \rangle$ for staggered fermions on the 9-cycled
and in instanton background
configurations. $\langle \bar \psi \psi \rangle$ in instanton background 
tracks the value of $\langle \bar \psi \psi \rangle$
measured on the 9-cycled configurations quite closely, down to small
quark mass.
It appears that the instantons,
present in equilibrium gauge field configurations of the QCD vacuum
generated
by Monte Carlo,
are breaking chiral symmetry by themselves.
This effect is  a cornerstone of
instanton-liquid models of hadron structure.

If  $\langle \bar \psi \psi \rangle$ is nonzero, one expects the spectrum
contains a would-be Goldstone boson (the pion) in addition to
massive quarks and (possibly) other resonances.
To test this hypothesis,  we fit the pseudoscalar correlator to two terms:
a pure hyperbolic cosine (a pole in the frequency plane), plus
a $q \bar q$ branch cut, with the quark mass as a parameter. We used
the
analytic expression for the branch cut
(expressed as a momentum and frequency mode sum),
 with the appropriate boundary conditions
and source used in the simulations (computed simply by measuring the
correlators on trivial background gauge configurations).

\begin{figure}
\centerline{\ewxy{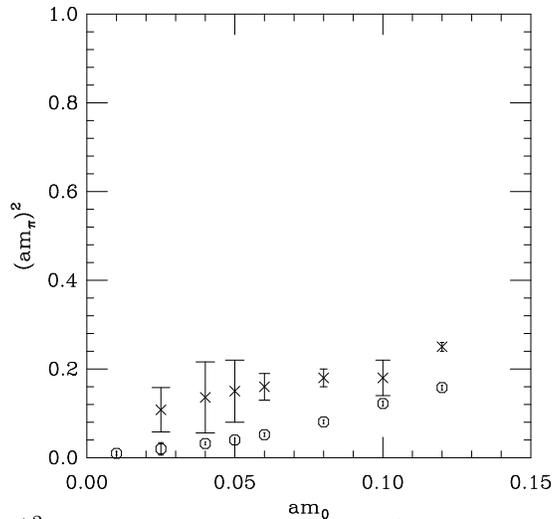}{80mm}
}
\caption{$(am_\pi)^2$ vs. $m_0$ the bare quark mass
for  staggered fermions in background instanton configurations,
with parallel (crosses) and randomly rotated instantons
  (octagons).}
\label{fig:pi2i}
\end{figure}

Since we have only 30 instanton configurations, the fits for the bound state
masses are quite poor. 
The signal in the oriented instanton background is quite noisy and the fit
masses are not well determined. 
  The resulting (squared) pion mass is shown in Fig. \ref{fig:pi2i}.
It is quite small  and appears to be decreasing
 as the bare mass vanishes. We cannot tell if it vanishes at zero
 bare quark mass.
The signal in the random instanton background is much larger and more
stable. We clearly see a light mass which decreases towards zero
as the quark mass vanishes.

The quark mass in the randomly oriented instanton
background is also   determined by the fit.
The  free correlator depends weakly (quadratically) on the
precise value of the quark mass when it is very small.
 At $am_0=0.10$ we
fit $am_q=0.109(2)$, 
0.105(3) at $am_0=0.08$, and
 0.08(4) at $am_0=0.04$.
The quark mass is less than half the rho mass--and less than half the
pion mass-- (as measured on
the 9-cycled lattices).

Are there other bound states? Only for the randomly rotated 
instanton configurations could we see a convincing pseudoscalar,
and so we restricted our analysis to that data set.  We again attempted
to fit to a single resonance plus a $q\bar q$ continuum.
Of course, this assumption is questionable.

Even in the usual
(confined) case, the other meson channels
 are dominated by two relatively light
 particles of opposite parity. We are attempting to fit the correlator
in the instanton background field to a free quark continuum (which
oscillates in time) and only a single resonance.
This may be incorrect. The correlator might have contributions from
two opposite parity bound states and a $q \bar q$ continuum, or no
continuum, or no resonances.
 So we may just be fitting the effects of
a large number of excited states, or of interactions between free quarks,
to a single exponential, and (incorrectly) interpreting our results
as evidence for a resonance.
The fits are not of high quality and our results should not be taken
too seriously:  In the ``SC'' channel (saturated by the $\pi_2$ and scalar
mesons in the confined phase) we saw a light bound state whose mass
roughly tracked the mass of the (presumed) pion resonance in the
 pseudoscalar channel, falling from about $am(\pi_2)=0.44$ at $am_0=0.12$
to about $am(\pi_2)=0.18$ at $am_0=0.04$.
In the PV and VT channels, a state with a mass 0.7-0.6 appears, in addition to
the free $q\bar q$ continuum.  This state has about the same mass
as the vector meson in the confined system, at an equivalent pseudoscalar mass.
However, the dominant feature of all these channels is still the
free $q\bar q$ continuum, with fitted quark masses of $am_q=0.1$ or lower.
The energy of the free $q\bar q$ continuum is always lower than the mass
of any (presumed) resonance.

Our work shows that instantons have broken chiral symmetry and
probably that a light pion has been generated.  However, instantons
 do not confine, and the properties of quarks propagating in instanton
background field are quite different from those of the full theory.

All this work is done at large lattice spacing of about $0.14$ $fm$.
Our work \cite{INST} has shown that the typical size of instantons in
SU(2) gauge theory is about $\rho \simeq 0.2$ $fm$, and it may well be that
staggered fermions do not couple well to such small (in units of the lattice
spacing) structures.  Thus it would be interesting to repeat this
work at smaller lattice spacing. Does the mass scale of the
free quark continuum ever rise above the mass scale of any
hadronic bound states?

It would  also be very interesting to repeat these studies for $SU(3)$ gauge
theory. However, the cycling transformation is simply too expensive to
use for large volumes in $SU(3)$.
It may be that it is possible to replace cycling by some kind of local
averaging of the fields on the original lattice, tuned to preserve
medium-scale physics, including
topological structures larger than some minimum size.
This construction is presently under study.

\section*{Acknowledgements}
We would like to thank  the Colorado High Energy experimental
groups for allowing us to use their work stations.
This work was supported by the U.S. Department of Energy.

\newcommand{\PL}[3]{{Phys. Lett.} {\bf #1} {(19#2)} #3}
\newcommand{\PR}[3]{{Phys. Rev.} {\bf #1} {(19#2)}  #3}
\newcommand{\NP}[3]{{Nucl. Phys.} {\bf #1} {(19#2)} #3}
\newcommand{\PRL}[3]{{Phys. Rev. Lett.} {\bf #1} {(19#2)} #3}
\newcommand{\PREPC}[3]{{Phys. Rep.} {\bf #1} {(19#2)}  #3}
\newcommand{\ZPHYS}[3]{{Z. Phys.} {\bf #1} {(19#2)} #3}
\newcommand{\ANN}[3]{{Ann. Phys. (N.Y.)} {\bf #1} {(19#2)} #3}
\newcommand{\HELV}[3]{{Helv. Phys. Acta} {\bf #1} {(19#2)} #3}
\newcommand{\NC}[3]{{Nuovo Cim.} {\bf #1} {(19#2)} #3}
\newcommand{\CMP}[3]{{Comm. Math. Phys.} {\bf #1} {(19#2)} #3}
\newcommand{\REVMP}[3]{{Rev. Mod. Phys.} {\bf #1} {(19#2)} #3}
\newcommand{\ADD}[3]{{\hspace{.1truecm}}{\bf #1} {(19#2)} #3}
\newcommand{\PA}[3] {{Physica} {\bf #1} {(19#2)} #3}
\newcommand{\JE}[3] {{JETP} {\bf #1} {(19#2)} #3}
\newcommand{\FS}[3] {{Nucl. Phys.} {\bf #1}{[FS#2]} {(19#2)} #3}

\end{document}